\documentclass[usegraphicx,usenatbib]{mn2e}

\title[Pulsational and evolutionary analysis of BS\,Com]
      {Pulsational and evolutionary analysis of the double-mode RR Lyrae star BS\,Com}
\author[I. D\'ek\'any et al.]
       {
        I.~D\'ek\'any,$^1$\thanks{e-mail: dekany@konkoly.hu}
        G.~Kov\'acs,$^1$
        J.~Jurcsik,$^1$
        R.~Szab\'o,$^1$
        M.~V\'aradi,$^2$
        \'A.~S\'odor,$^1$
        \newauthor 
        K.~Posztob\'anyi,$^3$
        Zs.~Hurta,$^4$
        K.~Vida,$^4$
        N.~Vityi,$^4$
        and A.~Szing$^5$\\
        $^1$Konkoly Observatory, H-1525 Budapest, PO~Box 67, Hungary\\
        $^2$Observatoire de Gen\`eve, Universit\'e de Gen\`eve, CH-1290
            Sauverny, Switzerland\\
        $^3$AEKI, KFKI Atomic Energy Research Institute, Thermohydraulic Department,
            H-1525 Budapest 114, PO~Box 49, Hungary\\
        $^4$E\"otv\"os University, Department of Astronomy, H-1518 Budapest,
            PO~Box 32, Hungary\\
        $^5$University of Szeged, Dept. of Exp. Physics and Astron. Obs.,
            H-6720 Szeged, D\'om t\'er 9, Hungary
       }
\date{Released: will be inserted later}

\pagerange{\pageref{firstpage}--\pageref{lastpage}} \pubyear{2002}

\begin{document}

\label{firstpage}

\maketitle

\begin{abstract}

	We derive the basic physical parameters of the field double-mode
	RR~Lyrae star BS\,Com from its observed periods and the requirement
	of consistency between the pulsational and evolutionary constraints.
	By using the current solar-scaled horizontal branch evolutionary models of
	\citet{pietrinferni04apj} and our linear non-adiabatic purely
	radiative pulsational models, we get $M / M_{\odot} = 0.698 \pm 0.004$,
	$\log(L/L_{\odot}) = 1.712 \pm 0.005$, $T_{\rm eff} = 6840 \pm 14$\,K,
	[Fe/H]=$-1.67 \pm 0.01$, where the errors are standard deviations
	assuming uniform age distribution along the full range of uncertainty in age.
	The last two parameters are in a good agreement with the ones derived from the
	observed $BVI_{\mathrm{C}}$ colours and the updated ATLAS9 stellar atmosphere
	models. We get $T_{\rm eff}=6842 \pm 10$\,K, [Fe/H]=$-1.58 \pm 0.11$,
	where the errors are purely statistical ones.
	It is remarkable that the derived parameters are nearly independent
	of stellar age at early evolutionary stages.
	Later stages, corresponding to the evolution toward the asymptotic giant branch 
	are most probably excluded because the required high temperatures are less likely
	to satisfy the constraints posed by the colours. We also show that our
	conclusions are only weakly sensitive to nonlinear period shifts predicted
	by current hydrodynamical models.
	
\end{abstract}


\section{Introduction}\label{sect:introduction}
	
	Double-mode radial pulsators play a very significant role in the 
	study of pulsating stars. This is because we can determine all fundamental 
	stellar parameters just by using a few observed parameters 
	and relying on the rather solid theory of linear radial pulsation (assuming 
	of course, that the mode identification is correct -- a question 
	often raised for typically low-amplitude pulsators whereas rarely 
	bothered with in the case of high/moderate-amplitude ones). In a 
	series of papers \citep{kw99apj,kovacs2000smc,kovacs2000lmc}
	we investigated the consistency of the method
	by using the observed colours and assumed cluster metallicities in deriving
	distances for globular clusters and for the Magellanic Clouds.
	These distances have been confirmed by independent studies, for example
	by direct HST parallax measurements of \citet{vL07mnras} and
	\citet{benedict07aj}, by \emph{Hipparcos}-based $\delta$~Scuti PL-relation
	of \citet*{mcnamara07aj}, by mixtures of parallax and interferometric
	methods of \citet{fouque07aap} and by Baade-Wesselink
	analyses \citep{storm04aap,kovacs03mnras}.
	In yet another application, \citet{buchler07apj} put limits on the 
	metallicities of Magellanic Cloud beat Cepheids through stellar evolution constraints. 
	
	Double-mode radial pulsators are also viable for nonlinear hydrodynamical 
	modelling. Although early attempts have already shown that sustained 
	double-mode pulsation is possible with purely radiative hydrodynamical 
	codes \citep[at least for RR~Lyrae stars, see][]{kovacs93apj}, 
	the modelling was put on a better sounding physical ground only after 
	time-dependent convection was more carefully included
	(for Cepheids by \citealt{kollath98apjl}, for RR~Lyrae stars by
	\citealt{feuchtinger98aap} and \citealt*{szabo04aap}). 
	Although these models are capable of producing stable double-mode
	pulsation in the appropriate neighbourhood of the physical parameter
	space and the model light curves are in reasonable agreement with the 
	observed ones, star-by-star modelling and application of more stringent 
	observational constraints are still lacking.
	
	Large-scale surveys aimed at microlensing and variability studies have 
	already yielded a considerable increase in the number of known double-mode 
	variables. In particular, before 2000, only four
	double-mode RR~Lyrae (RRd) stars were identified in the Milky Way 
	(see \citealt{garcia-melendo97aj} and \citealt{kovacs01book} for additional 
	statistics). At the moment of this writing we have 26 RRd stars
	known in the Galactic field according to the literature, mostly
	due to the NSVS and ASAS full-sky variability surveys,
	and many more in the Galactic Bulge, due to the OGLE survey
	(\citealt{mizerski03aca}; \citealt{moskalik03aap} and \citealt*{pigulski03aca}). 
	
	The field RR Lyrae star BS\,Com ($\alpha = 13^{\rm h} 34^{\rm m} 39^{\rm s}$,
	$\delta = +24^{\circ} 16' 38''$) was originally put on the 
	observing list of the 60\,cm automated telescope of the Konkoly Observatory 
	(see \citealt{sodor07an} for details on the ongoing photometric survey),
	due to the suspected Blazhko 
	behaviour of this star reported by \citet{clementini95mnras}. While gathering data,
	the star was analysed by \citet{wils06ibvs}, using the NSVS database.
	He found that BS\,Com was an RRd star\footnote{We note however,
	that \citet{wils06ibvs} selected the $\sim\,1$~yr alias of the fundamental period.}. 
	
	This paper describes the results of the frequency analysis of our data 
	and the pulsational/evolutionary analysis. The result of this latter, 
	purely theoretical study will be confronted with additional constraints 
	posed by the effective temperature and metallicity obtained from our 
	multi-colour data. We show that the purely theoretical approach yields 
	stellar parameters in agreement with the colour data and can be 
	used for very accurate parameter estimation once we are able to pose 
	an age limit for the object.

\section{Photometry}\label{sect:photometry}

\subsection{Observations and data reduction}
\label{subsect:obs}

	We obtained $BV(RI)_C$ photometric observations of BS\,Com with
	the 60~cm automated telescope of the Konkoly Observatory, Budapest.
	The telescope is equipped with a Wright $750\times1100$ CCD providing a
	$17'\times24'$ field of view
	(detailed parameters are given by \citealt{bakos99}).
	The observations were carried out between 2005 March 1 and 
	2006 April 17 in 3 seasons, on a total number of 35 nights.
	Observational statistics are shown in Table~\ref{tab:obstat}.
	BS\,Com magnitudes were measured relative to BD\,+24~2598 by aperture 
	photometry\footnote{Photometric data are available 
	electronically on-line at CDS.}. 
	GSC\,00454-00454 and USNO-B1\,1143-0206728 served as check stars. 
	We used standard IRAF\footnote{IRAF is distributed by the 
	National Optical Astronomy Observatory, which is operated by 
	the Association of Universities for Research in Astronomy, Inc., 
	under cooperative agreement with the National Science Foundation.} 
	packages for data reduction procedures. Coefficients for linear 
	transformation of the instrumental data into standard magnitudes 
	were derived each year.
	
	BS\,Com was tied to the standard photometric system in 
	$B,V$ and $I_{\rm C}$ colours through the comparison star BD+24~2598.
	The magnitudes of BD\,+24~2598 were measured relative to HD\,117876 
	($V = 6$\hbox{$.\!\!^{\rm m}$}$082$, 
	$B-V = 0$\hbox{$.\!\!^{\rm m}$}$968$, 
	$V-R = 0$\hbox{$.\!\!^{\rm m}$}$522$, 
	$V-I = 1$\hbox{$.\!\!^{\rm m}$}$017$, 
	see \citealt{holtzman84aj}), 
	the closest standard star to BS\,Com. 
	The observations were carried out near two subsequent culminations, 
	under very good weather conditions.
	Because of the brightness of HD\,117876, a large fraction of the 
	incoming light had to be blocked in order to avoid saturation.
	We reduced the effective aperture of the telescope to the $\sim\,7 \%$ 
	of its original size by placing a circular diaphragm at the front 
	end of the telescope tube.
	Because of the $\sim\,24'$ angular distance between HD\,117876 and 
	BD+24\,2598, it was not possible to observe the stars in a common 
	field of view. In order to keep our photometric precision, the 
	two distinct fields were observed repeatedly. 
	We obtained 5, 10, and 9 frames in $B$, $V$, and $I_{\mathrm{C}}$, 
	respectively, for each of the two stars. 
	Data were corrected for atmospheric extinction. 
	The standard magnitudes and their formal statistical 
	errors are shown in Table~\ref{tab:mags}. For BS\,Com, magnitude
	averages are shown, calculated over the complete light 
	curve solution given in Sect.~\ref{subsect:fourier}.
	
	\begin{table}
	\caption{BS Com observational statistics}
	\label{tab:obstat}
	\begin{tabular}{cccccc}
	\hline
	date & nights & \multicolumn{4}{c}{data points}\\
	$(\mathrm{JD} - 2450000)$ & & $B$ & $V$ & $R_{\mathrm{C}}$ & $I_{\mathrm{C}}$\\
	\hline
	$3431$ -- $3467$ & $19$ & $568$ & $566$ & $561$ & $543$\\
	$3744$ -- $3762$ & $7$ & $0$ & $148$ & $149$ & $149$\\
	$3824$ -- $3843$ & $9$ & $0$ & $256$ & $214$ & $243$\\
	\hline
	total: & $35$ & $568$  & $970$ & $924$ & $935$\\
	\hline
	\end{tabular}
	\end{table}

	\begin{table}
	\caption{Standard $BVI_{\mathrm{C}}$ magnitudes}
	\begin{flushleft}
	\label{tab:mags}
	\begin{tabular}{lr@{.}lr@{.}lr@{.}l}
	\hline
	Star & 
	\multicolumn{2}{c}{$V$} & 
	\multicolumn{2}{c}{$B-V$} & 
	\multicolumn{2}{c}{$V-I_{\mathrm{C}}$}\\
	\hline
	BD+24~2598 & 10&608 & 0&603 & 0&662\\
	BS\,Com & 12&755 & 0&286 & 0&433 \\
	errors: & $\pm 0$&0015 & $\pm 0$&0016 & $\pm 0$&0032\\
	\hline
	\end{tabular}
	\end{flushleft}
	\begin{flushleft}   
	{\footnotesize \underline {Notes:}
	Magnitude averages are given. 
	The errors are standard deviations of the 
	mean differences between BD\,+24~2598 and HD\,117876.}
	\end{flushleft}
	\end{table}

\subsection{Fourier analysis: the double-mode solution}
\label{subsect:fourier}

	We performed a standard frequency analysis on the $V$ 
	data by discrete Fourier-transformation. 
	We derived $P_0 = 0.48791$\,d and $P_1 = 0.36307$\,d for 
	the periods of the fundamental and the first overtone modes, 
	respectively. 
	Besides the frequencies of the two pulsation modes, 
	$5$ harmonics and $8$ linear combinations of the 
	frequencies corresponding to the above periods were 
	identified in the amplitude spectra. In order to 
	increase the detection probability of potentially 
	existing additional components, we applied the spectrum 
	averaging method described by \citet{nagy06aa}. 
	Due to the averaging of the frequency spectra in 
	the $V(RI)_C$ colours, the noise of the residual spectra 
	decreased by a factor of $\sim\,\sqrt{3}$, down to the $\sim\,2$\,mmag level. 
	Except for the above two components, their harmonics, 
	and linear combinations, no additional components 
	were found. More details on the frequency analysis 
	are given by \citet{dekany07an}. 
	Our 15-term Fourier-sum fit gives a full description 
	of the light curves within the observational accuracy. 
	The Fourier-parameters and their errors are given in 
	Table~\ref{tab:fourier}.
	
\section{
         Photometric abundance and temperature from 
         $\bld{B, V, I_{\mathrm{C}}}$ colours
        }
\label{sect:abundances}

	Knowledge of the metallicity and temperature is 
	very important in the determination of the physical 
	parameters of double-mode stars (\citealt{cox91apj}; 
	\citealt{kovacs92aap} and \citealt{kovacs2000lmc}).
	There are two former estimates of the metallicity 
	of BS\,Com. \citet{smith90pasp} obtained two measurements 
	of $\Delta S = 6.0$ and $8.2$. The second value was taken 
	at lower brightness than the first one, so this, being 
	closer to minimum light yields $\mathrm{[Fe/H]} = -1.53$ 
	on the scale of \mbox{\citet{j95}}.
	\citet{clementini95mnras} acquired one spectrum of BS\,Com
	with a resolution of 1.4\,\AA~and a signal-to-noise
	ratio of $\sim\,20$. They derived a very low abundance
	of ${\rm [Fe/H]} \sim\,-2.0$ from the equivalent width
	of the Calcium K-line. As explained by \citet{clementini95mnras},
	they took the spectrum `at minimum light',
	according to their photometric $V$ time series 
	consisting of $97$ points. However, taking into 
	consideration the double-mode behaviour of BS\,Com
	yet unknown at that time, the real pulsational phase 
	of the aforementioned observations is ambiguous. 
	(Unfortunately, the pulsational parameters of 
	BS\,Com derived in this paper do not enable us 
	to predict the light curve accurately enough at the time 
	of these observations.) 
	Therefore, we decided to make an independent 
	estimation of the metallicity and also the effective 
	temperature of BS\,Com based on our photometric data.

	The idea of deriving heavy element abundances from magnitudes 
	measured in proper photometric bands is not new.
	In his early study, \cite{sturch} used $UBV$ magnitudes around
	minimum light to estimate the line blanketing of RRab stars. 
	In a similar application, \cite{lub} employed the reddening-free
	Walraven [$B-L$] colour index for metallicity determination.
	Also, we refer to a more recent paper by \citet*{twarog03aj},
	who obtained more precise abundances for open clusters using 
	intermediate-band photometry.
	
	The method represents the most economical approach in 
	terms of getting a first order estimate on [M/H].
	In a completely formal approach, the method of photometric 
	abundance determination comes from the following idea. 
	In the static stellar atmosphere computations the models 
	are determined by fixing basic input parameters, such as 
	chemical composition, convective parameters 
	(mixing length over atmospheric scale height, 
	turbulent velocity, overshooting, etc.), 
	gravitational acceleration $g$ and effective 
	temperature $T_\mathrm{eff}$. If we fix the 
	admittedly poorly known convective parameters 
	and assume that solar-type abundance 
	distribution is, in general, a good approximation 
	of the true distribution, we are left with three parameters 
	only: overall metal abundance [M/H], $\log{g}$, and $T_\mathrm{eff}$. 
	Then, it is obvious that once we have an independent 
	estimate on $\log{g}$, then we can compute both 
	$T_\mathrm{eff}$ and [M/H] from two colour indices, 
	assuming that they depend differently on the above 
	two parameters. Because of the availability of the 
	average $B, V, I_{\mathrm{C}}$ magnitudes of BS\,Com 
	and because the $B-V$ colour index relatively 
	strongly depends on [M/H] \citep[see e.g.][]{kw99apj}, 
	whereas $V-I_{\mathrm{C}}$ is mostly sensitive to 
	$T_\mathrm{eff}$, it is useful to examine the possibility 
	of determining [M/H] and $T_{\rm eff}$ through the 
	$B-V$ and $V-I_{\mathrm{C}}$ colour indices.
	
	Because of the poor representation of the $B-V$, [M/H], $\log{g}$, 
	and $T_\mathrm{eff}$ dependence by a single linear or low-order 
	polynomial formula, we decided to use directly the grids available 
	for stellar atmosphere models. We employ quadratic interpolation 
	in order to get accurate model colour indices for each set of physical 
	parameters. More precisely, having the value of $\log{g}$ fixed, 
	we scan the parameter range in $B-V$, $V-I_{\mathrm{C}}$ and find 
	the best [M/H], $\log{T_\mathrm{eff}}$ parameters that minimise the 
	following function:
	
	{\setlength\arraycolsep{2pt}
	\begin{eqnarray}
	\label{discriminator}
	\mathcal{D} & = &
	\left[ \log{T_\mathrm{eff}} (B-V) - 
	       \log{T_\mathrm{eff}} (V-I_{\mathrm{C}}) \right]^2
	\nonumber \\
	& & + \alpha^2_{B-V}\left[ (B-V)_{\rm obs} - (B-V) \right]^2
	\nonumber \\
	& & + \alpha^2_{V-I_{\mathrm{C}}}\left[ (V-I_{\mathrm{C}})_{\rm obs} - 
	(V-I_{\mathrm{C}}) \right]^2~,
	\end{eqnarray}}

	\noindent
	where $T_\mathrm{eff}(B-V)$ and $T_\mathrm{eff}(V-I_{\mathrm{C}})$ 
	are the $T_\mathrm{eff}$ values at the respective colour indices, 
	the subscript `obs' means the observed values, whereas the scanned 
	values are denoted without subscripts. The weights $\alpha_{B-V}$ 
	and $\alpha_{V-I_{\mathrm{C}}}$ are set equal to $0.33$ and $0.25$, 
	respectively, in order to account for the proportionality of 
	$\log{T_\mathrm{eff}}$ to $0.33(B-V)$ and $0.25(V-I_{\mathrm{C}})$ 
	in formulae obtained by linear regressions \citep[see][]{kw99apj}. 
	The introduction of the last two terms in $\mathcal{D}$ is necessary, 
	because of the high sensitivity of the method for observational noise.
		
	We employ the stellar atmosphere models of 
	\citet[][see also \citealt*{castelli97aap}]{castelli99aap} 
	computed with the opacity distribution functions based on the solar 
	abundances of \citet{grevesse98ssr} (grids quoted as `ODFNEW' at 
	Kurucz's web site\footnote{http://kurucz.harvard.edu}). 
	These new grids differ from the previous ones as described in 
	\citet{pietrinferni04apj}. All models have a microturbulent velocity 
	of $2\,\mathrm{km\,s^{-1}}$, a convective mixing length parameter 
	of $1.25$, scaled solar heavy element distribution and 
	no convective overshooting. For the computation 
	of $\log{g}$ we used a similar formula to that in 
	\citet{kovacs2000smc}, adapted to the appropriate
	parameter regime of RR Lyrae stars:
	\begin{eqnarray}
	\label{logg}
	\log{g} = 2.47 - 1.23\log{P_0}~.
	\end{eqnarray}
	
	We note two possible sources of ambiguity when using 
	Eq.~(\ref{discriminator}) for the determination of 
	$T_{\rm eff}$ and [M/H]. First, the zero points of the 
	various colour\,---\,$T_{\rm eff}$ calibrations
	(e.g. infrared flux method, IRFM versus calibrations 
	based on Vega or Sun-like stars) may be 
	slightly different. In earlier applications we used zero points 
	defined by IRFM \citep[see e.g.][]{kovacs2000smc}. For pulsating stars,  
	there is also a problem of the difference between the static and 
	pulsation-phase-averaged colours \citep*{bono95apjs}. Since the 
	sizes of these effects are still debatable, we opted to use the 
	stellar atmosphere data by \citet{castelli99aap} without applying any 
	corrections to the magnitude-averaged colours. As we shall see in the 
	subsequent test, the derived metallicities are in a good agreement with 
	the spectroscopic ones, indicating that the theoretical colours are fairly accurate.

	\begin{figure}
	\includegraphics[width=84mm]{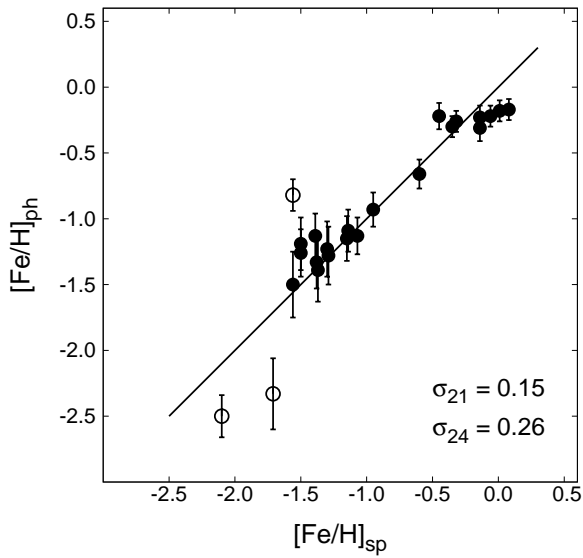}
	\caption{
	         [Fe/H]$_{\mathrm{ph}}$ values computed from the 
	         $BVI_{\mathrm{C}}$ colours plotted against the 
	         spectroscopic [Fe/H]$_{\mathrm{sp}}$ values taken 
	         from the literature for 24 RRab stars. Error bars 
	         denote standard deviations for each star, obtained by adding 
	         uncorrelated Gaussian 
	         random noise with $\sigma = 0$\hbox{$.\!\!^{\rm m}$}$005$ 
	         to the observed $B-V$, $V-I_{\mathrm{C}}$ colour indices.
	         Open circles show the outlying stars X\,Ari, SS\,Leo, and 
	         VY\,Ser. 
	        }
	\label{fig:bvifeh}
	\end{figure}

	To test the reliability of the method we collected 24 fundamental 
	mode RR Lyrae (RRab) stars from the literature with known periods, 
	$B-V$ and $V-I_{\mathrm{C}}$ average colour indices, and spectroscopic 
	[Fe/H] values\footnote{For solar-scaled heavy element distribution 
	we have $[\mathrm{M/H}] = [\mathrm{Fe/H}]$.}. Details on this test sample are 
	given in Appendix~\ref{appendix:teststars}. In Fig.~\ref{fig:bvifeh} 
	we show the derived [Fe/H]$_{\mathrm{ph}}$ values versus the 
	[Fe/H]$_{\mathrm{sp}}$ spectroscopic metallicities of the test stars. 
	By the exclusion of the three stars X\,Ari, SS\,Leo, and VY\,Ser with
	discrepant photometric metallicities (open circles in Fig.~1), the standard
	deviation of the fit decreases from $0.26$~dex to $0.15$~dex.
	We consider the latter as the $1\,\sigma$ error of the method.
	Estimates of the errors of the individual data points were computed by 
	adding Gaussian random noise of $\sigma = 0$\hbox{$.\!\!^{\rm m}$}$005$ 
	to the input colour indices. 	
	The error bars in Fig.~\ref{fig:bvifeh} indicate that the sensitivity 
	of the method to observational errors increases with decreasing 
	[Fe/H] (see also Kov\'acs 2008 for more details).

	We estimated the metallicity and effective temperature of BS\,Com 
	by using $\log g = 2.85$ from Eq.~(\ref{logg}) and the average 
	$B-V$ and $V-I_{\mathrm C}$ values given in Table~\ref{tab:mags}. 
	The amount of reddening at the high Galactic latitude of BS\,Com 
	($l = 19\fdg00$, $b = +79\fdg86$) is expected to be low. 
	Indeed, according to \citet*{schlegel98apj}, $\mathrm{E}(B-V) = 0.014$. 
	This gives a maximum value for the real reddening of the star, 
	and therefore poses minimum values to its intrinsic colours. 
	The estimated metallicities and effective temperatures of BS\,Com 
	with different possible reddening values are shown in 
	Table~\ref{tab:bsctfeh}. Formal statistical errors are $10$\,K 
	and $0.11$~dex, computed the same way as for the test objects, 
	but using the photometric errors listed in Table~\ref{tab:mags}.
			
\section{Pulsation models}
	
	We have generated linear non-adiabatic (LNA), fully radiative stellar 
	models to derive the fundamental physical parameters of BS\,Com. The 
	pulsation code is basically identical to the one described by 
	\citet{buchler90}, originally developed by \citet{stellingwerf75} 
	\citep[see also][]{castor71}.
	As in \citet{kovacs06mmsai}, we used the Rosseland mean opacity values 
	taken from \citet{iglesias96apj} and \citet{alexander94} interpolated to the 
	appropriate chemical composition. The heavy element distribution 
	is solar-scaled, as given by \citet{grevesse91}. The stellar 
	envelopes were densely sampled with 500 mass shells down to the 
	inner radius and mass values, where
	$Q_{\rm in}\equiv (R_{\rm in}/R_{\rm surf})(M_{\rm in}/M_{\rm surf})=0.05$, 
	with $R_{\rm surf}$ and $M_{\rm surf}$ values at the outermost 
	grid point. This zoning resulted in a deep envelope down to 
	$(6-7)\cdot10^6$\,K and ensured a sufficient (better than $10^{-4}$~d) 
	stability in the periods (as compared to shallower envelopes of 
	$Q_{\rm in} = 0.10$).
	
	\begin{table}
	\caption{
	         Reddening dependence of the photometric [Fe/H] and 
	         $T_{\mathrm{eff}}$ values of BS\,Com.
	        }
	\begin{flushleft}       
	\label{tab:bsctfeh}
	\begin{tabular}{llc}
	\hline
	$\mathrm{E}(B-V)$ & [Fe/H] & $T_{\mathrm{eff}}$ \\
	\hline
	$0.000$   & $-1.47$ & $6774$~K \\
	$0.005$   & $-1.51$ & $6798$~K \\
	$0.010$   & $-1.54$ & $6823$~K \\
	$0.014^*$ & $-1.58$ & $6842$~K \\
	\hline
	\end{tabular}
	\end{flushleft}
	{\footnotesize $^{*}$ \citet{schlegel98apj}}
	\end{table}

	We covered the metallicity and temperature ranges well beyond the ones 
	derived from the $BVI_{\rm C}$ colours in Sect.~\ref{sect:abundances}. 
	By scanning these ranges and matching the observed 
	periods, we obtained unique solutions for the possible physical parameters 
	of the star. These solutions are listed in Table~\ref{tab:lnagrid}.
	To be more compatible with the evolutionary models derived by
	\citet{pietrinferni04apj}, with the change of the overall metal content $Z$,
	we have also changed the hydrogen abundance $X$. We note that our best 
	estimated [Fe/H] and $T_{\rm eff}$ values are closest to the 
	item at $Z=0.0006$ and $T_{\rm eff}=6850$\,K. It is noticeable, 
	how sensitive the solution is to the metal content. For example, 
	by changing [Fe/H] by $\sim \,0.2$\,dex, the mass changes by 
	$0.1$\,--\,$0.15$\,$M_{\odot}$. However, a $200$\,K change in $T_{\rm eff}$ 
	causes a change in the mass less than $0.06$\,$M_{\odot}$. We 
	also draw attention to the remarkable stability of the gravity 
	and density. While the [Fe/H] change mentioned above causes  
	$\sim\,15$~\% change in mass, the same figure is less than 
	$0.8$\,\% and $0.3$\,\% in $\log g$ and $\log \rho$, respectively. 
	The effect of temperature change is even smaller.
	It is also important to note that all models listed 
	in the table are linearly excited in their fundamental and first 
	overtone modes. Therefore, they are viable for sustained double-mode 
	pulsations. More detailed tests by direct nonlinear hydrodynamic 
	simulations are beyond the scope of this paper (see, however Section
	\ref{sect:esps}).

	\begin{table}
	\caption{Pulsation models matching the periods of BS\,Com}
	\label{tab:lnagrid}
	\begin{flushleft}                                                     
	\begin{tabular}{cccccc}                                         
	\hline\hline
	$T_{\rm eff}$ & $M$ & $\log L$ & $R$ & $\log g$ & $\log \rho$  \\
	\hline
	&&\multicolumn{2}{c}{$X=0.7547$} & {$Z=0.0003$}\\
	\hline
		6700 & 0.639 & 1.647 & 4.940 & 2.856 & $-2.127$ \\
		6750 & 0.646 & 1.664 & 4.964 & 2.857 & $-2.128$ \\ 
		6800 & 0.656 & 1.682 & 4.993 & 2.858 & $-2.130$ \\
		6850 & 0.663 & 1.699 & 5.018 & 2.858 & $-2.131$ \\
		6900 & 0.673 & 1.717 & 5.049 & 2.860 & $-2.133$ \\
	\hline
	&&\multicolumn{2}{c}{$X=0.7546$} & {$Z=0.0004$}\\
	\hline
		6700 & 0.669 & 1.661 & 5.021 & 2.861 & $-2.128$ \\
		6750 & 0.678 & 1.679 & 5.050 & 2.862 & $-2.130$ \\
		6800 & 0.688 & 1.697 & 5.080 & 2.863 & $-2.131$ \\
		6850 & 0.698 & 1.715 & 5.111 & 2.865 & $-2.133$ \\
		6900 & 0.709 & 1.733 & 5.143 & 2.866 & $-2.134$ \\
	\hline
	&&\multicolumn{2}{c}{$X=0.7534$} & {$Z=0.0006$}\\
	\hline
		6700 & 0.732 & 1.690 & 5.188 & 2.872 & $-2.132$ \\
		6750 & 0.743 & 1.708 & 5.221 & 2.873 & $-2.133$ \\
		6800 & 0.754 & 1.726 & 5.253 & 2.875 & $-2.135$ \\
		6850 & 0.768 & 1.745 & 5.291 & 2.876 & $-2.137$ \\
		6900 & 0.780 & 1.763 & 5.324 & 2.877 & $-2.138$ \\
	\hline
	&&\multicolumn{2}{c}{$X=0.7530$} & {$Z=0.0010$}\\
	\hline
		6700 & 0.872 & 1.745 & 5.530 & 2.893 & $-2.139$ \\
		6750 & 0.885 & 1.763 & 5.563 & 2.894 & $-2.140$ \\
		6800 & 0.901 & 1.782 & 5.603 & 2.896 & $-2.142$ \\
		6850 & 0.914 & 1.800 & 5.637 & 2.897 & $-2.143$ \\
		6900 & 0.929 & 1.818 & 5.672 & 2.898 & $-2.145$ \\
	\hline                                                                
	\end{tabular}                                                         
	\end{flushleft}
	\begin{flushleft}
	{\footnotesize \underline {Notes:} 
	$T_{\rm eff}$ is given in [K], $M$, $\log L$ and $R$ are measured in 
	solar units, $g$ and $\rho$ are in [CGS]. Observed periods match 
	within $10^{-5}$~d.}
	\end{flushleft}
	\end{table}

\section{Combination of the pulsational and evolutionary models}
\label{sect:esps}

	Stellar evolutionary models play a crucial role in 
	understanding basic stellar physics and developing more coherent 
	models. For example, the `Cepheid mass discrepancy' problem before 
	1992 appeared as a basic confrontation between the `standard' 
	evolutionary and pulsational theories. A desperate search for the 
	clue of this discrepancy led Norman Simon to a plea for the re-visitation 
	of the heavy element opacities \citep{simon82apjl}, that has resulted 
	finally in a success in 1992 (\citealt{rogers92apjs}, see also
	\citealt{seaton94mnras}). In another 
	application of the evolutionary results, in the current search for 
	extrasolar planets, a generally used method for determining the 
	physical parameters of the host star is matching stellar isochrones 
	in the $\rho_{\rm star}$ -- $T_{\rm eff}$ plane to the observed 
	values (assuming that $T_{\rm eff}$ and [Fe/H] are reliably determined 
	from e.g. spectroscopy, see \citealt{sozzetti07apj}).
	
	From a completely formal point of view, the method of parameter
	determination by the combination of evolutionary and pulsational models
	is equivalent to the following problem. At a certain pair of periods,
	the LNA pulsational model grid (Table~\ref{tab:lnagrid}) gives stellar
	mass and luminosity as a function of temperature and metallicity.
	Moreover, HB evolutionary tracks of \citet{pietrinferni04apj}
	also give stellar mass and
	luminosity for arbitrary temperature, metallicity, and age values.
	Therefore, we have the following two vector functions sampled in discrete points:
	\begin{eqnarray}
	\label{esps}
	\nonumber &{\mathbf E}:&~(\mathrm{[Fe/H]}, T_{\rm eff}, t) \rightarrow (M, L)\\
	\nonumber &{\mathbf P}:&~(\mathrm{[Fe/H]}, T_{\rm eff}, P_0, P_1) \rightarrow (M, L)~,
	\end{eqnarray}
	where ${\mathbf E}$ stands for the evolutionary models, ${\mathbf P}$
	denotes the pulsational models, and $t$ is the time elapsed from the
	start of the core helium burning at the zero age horizontal branch (ZAHB).
	If we assume consistency between the two kinds of models,
	then the subspace of stellar parameters that fulfil the constraints
	posed by both models is given by the intersection of the two vector 
	functions $({\mathbf E} = {\mathbf P})$.
	This will be (at the most) a 1D subspace, i.e. a set of solutions
	that has age $t$ as a free parameter.
	The [Fe/H] and $T_{\rm eff}$ ranges corresponding to our multitude of solutions
	will then be compared with the ones obtained from the photometry in
	Section~\ref{sect:abundances}. The method is implemented as follows.

	\begin{figure}
	\label{fig:onesolution}
	\centering
	\includegraphics[width=84mm]{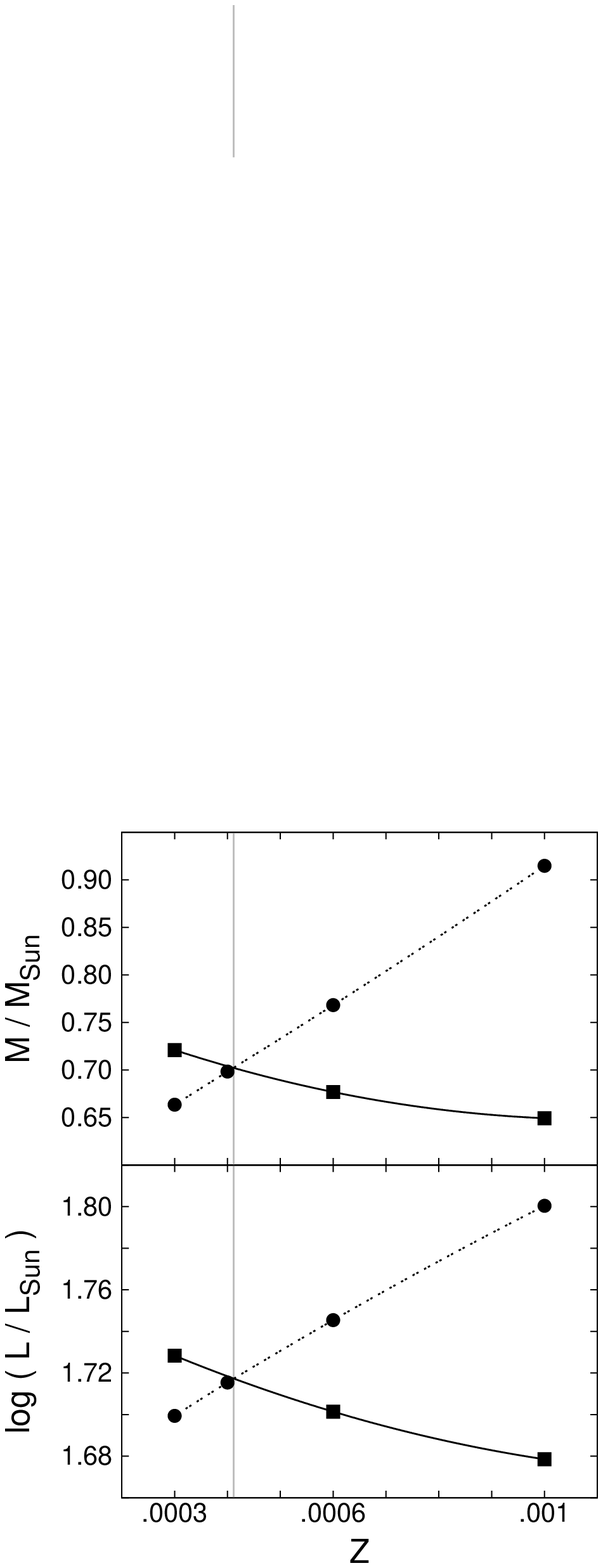}
	\caption{Stellar mass and luminosity as functions of the 
		 metallicity. Age and temperature are fixed at
		 $\log(t[\mathrm{yr}])= 7.424$
		 and $T_{\rm eff} = 6851$\,K. Continuous 
		 and dashed lines show the quadratic interpolants in 
		 Z for the evolutionary (squares) and pulsational
		 (circles) models. For better visibility, the
		 intersections are connected by a vertical grey line.
		}
	\end{figure}

	\begin{description}
	\item[(a)]
	We construct isochrones from the solar-scaled horizontal branch models of
	\citet{pietrinferni04apj}. Physical ingredients and details of the model
	building are described in their paper and references therein.
	Here we note only some relevant features. All models are canonical in the sense
	that they have been calculated without atomic diffusion and convective
	overshooting. They include up-to-date input physics, most importantly,
	they use recent and accurate results for plasma-neutrino processes and
	electron conduction opacities which lead to more accurate He core masses
	at evolutionary stages after the core helium flash. And last, but not least,
	their HB models have a fine mass spacing, particularly
	suitable for investigating stellar parameters in the instability strip.

	\item[(b)]
	From the pulsational solution (PS) we get `near straight lines' 
	for any fixed $T_{\rm eff}$ in the $Z~\rightarrow~(M, \log L)$
	planes (hereafter ML planes).
	
	\item[(c)]
	For a given age we can search for the evolutionary solution (ES)
	by using the above fixed $T_{\rm eff}$. These solutions yield 
	another set of `near straight lines' in the ML planes.
	
	\item[(d)]
	The ES and PS lines may or may not intersect in the appropriate
	$Z$ range in the ML planes.
	If they do, then we have two $Z$ values at which intersection occurs.
	Obviously, the consistent $\mathrm{ES} = \mathrm{PS}$ solution
	should yield the same $Z$.
	Since this is generally not the case, the best solution is searched 
	for by minimising the difference between the two metallicities:
	\begin{eqnarray}
	\label{eqn:deltaz}
	\Delta Z & = & |\log Z_M - \log Z_L|\hskip 2mm ,
	\end{eqnarray}
	where $Z_M$ and $Z_L$ are the metallicities obtained in the M and 
	L planes, respectively. To illustrate the solution obtained at 
	a given age and temperature, we show the PS and ES curves and their 
	intersections in Fig.~2. It is seen that these curves are indeed 
	near linear. Therefore, we can employ quadratic interpolation 
	to compute the positions of the intersections. 
	
	\item[(e)]
	By changing the age, with the above method we can map the solution 
	as a function of the evolutionary stage. As it follows from the above 
	construction, at each stage we have different stellar parameters 
	that yield the same periods and satisfy evolutionary constraints. 
	The grey-scale plot of ${\Delta Z}$ on the $\log(t)$\,---\,$T_{\rm eff}$
	plane is shown in Fig.~3.
	It is remarkable how much the solution 
	is limited in a narrow $T_{\rm eff}$ regime. It is also observable 
	that we get solutions (i.e. ${\Delta Z}<10^{-4}$) at nearly all 
	ages up to $\log(t)=7.77$ ($\approx60$~Myr).
	We could, in principle, fix the 
	age by using very accurately (and independently) determined 
	$T_{\rm eff}$. However, the precision of the currently available 
	methods for temperature estimation is insufficient for posing a strong
	constraint on the age.
	\end{description}
	
	\begin{table}
	\label{tab:bscparam}
	\caption{Physical parameters of BS\,Com derived in this paper}
	\begin{flushleft}
	\begin{tabular}{rrcc}
	\hline
	\multicolumn{1}{c}{[Fe/H]} & \multicolumn{1}{c}{$T_{\rm eff}$}[K] & $M/M_{\odot}$ & $\log(L/L_{\odot})$ \\
	\hline
	$-1.67 \pm 0.01$ & $6840 \pm 14$ & $0.698 \pm 0.004$ & $1.712 \pm 0.005$ \\
	$-1.58 \pm 0.11$ & $6842 \pm 10$ & & \\
	\hline
	\end{tabular}
	\end{flushleft}
	\begin{flushleft}
	{
	 \footnotesize
	 \underline {Note:}
	 Errors for the ES/PS solution (first line) are the standard deviations 
	 of the various age-dependent solutions in the full age interval of $60$~Myr. 
	 Items derived from the $BVI_{\rm C}$ colours (second line) are taken from 
	 Table~\ref{tab:bsctfeh}, errors are formal statistical ones.
	}
	\end{flushleft}
	\end{table}

	The ridge line in Fig.~3 selects the best fitting models. 
	The age dependencies of the corresponding $\log L$, $M$, and [Fe/H] 
	are shown in Fig.~4. The small range of stellar parameters allowed 
	by the solution is remarkable. The stronger topological change 
	occurs at $\log t=7.6$ ($\approx 40$~Myr) just after the 
	blueward loop where evolution starts toward the asymptotic giant branch. 
	Considerably larger ages seem to be less probable because the associated high 
	temperatures can be excluded by the ones derived from the colours.   
	From the above result we can estimate the parameters 
	of BS\,Com. Assuming equal probability of ages in the range plotted,
	we can compute the distributions of the various physical parameters.
	Although these distributions are not symmetric around their most
	probable values, for the present case this effect is of secondary 
	importance and we compute the standard deviations around the averages 
	as a first approximation of the errors introduced by the effect of 
	age uncertainty. Table~\ref{tab:bscparam} displays the so-derived 
	stellar parameters, which are compared with the independently estimated 
	$T_{\rm eff}$ and [Fe/H] values from Table~\ref{tab:bsctfeh}.
	It is reassuring that there is a comfortable overlap between the
	two sets of parameters, indicating a good level of consistency
	among the very different ingredients of the above parameter determination.
	
	\begin{figure}
	\label{fig:grayscale}
	\centering
	\includegraphics[width=84mm]{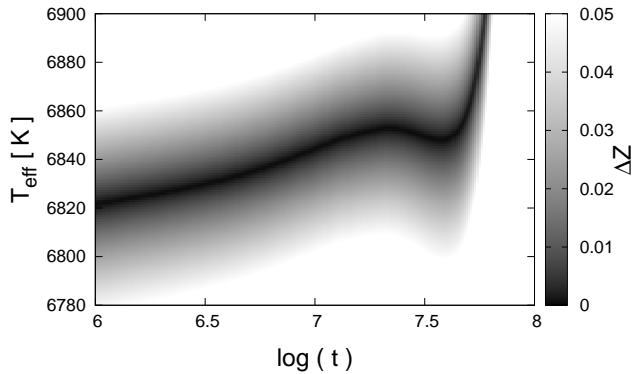}
	\caption{
	         Grey-scale plot of the metallicity difference $\Delta Z$ 
		 [see Eq.~(\ref{eqn:deltaz})] on the age\,---\,$T_{\rm eff}$ plane.
		}
	\end{figure}

	\begin{figure}
	\label{fig:ridge}
	\centering
	\includegraphics[width=84mm]{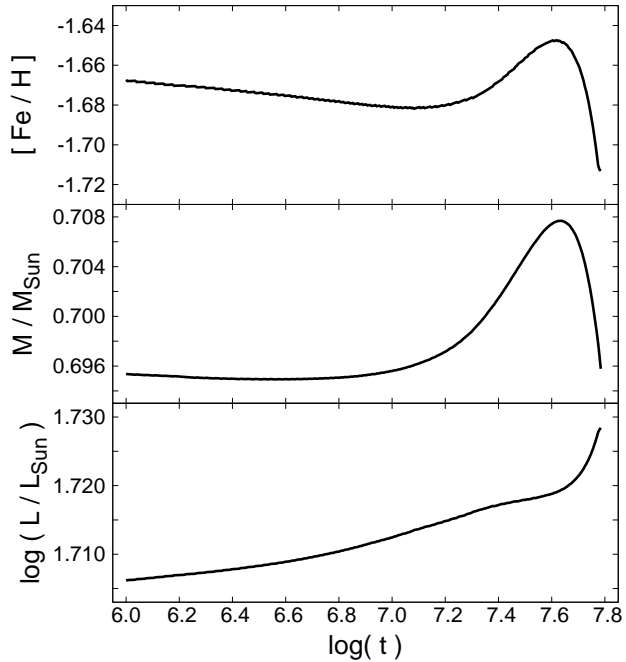}
	\caption{
	         Age dependence of the derived stellar parameters for BS\,Com.
	        }
	\end{figure}
	
	The employment of the LNA pulsation models in the above solution presumes 
	that the theoretical periods are applicable to the observed ones that 
	are known to be inherently nonlinear due to the longevity of the 
	observable double-mode state. Since our method strongly relies on 
	this assumption, it is important to examine if the currently available 
	nonlinear models predict some considerable change in the asymptotic 
	nonlinear periods with respect to the LNA periods. We note that this 
	problem is also present in any asteroseismological investigation if 
	the multimode state is the result of some nonlinear phenomenon. 
	However, nonlinear modelling is usually not easy to attempt since 
	most of the multimode pulsators are of non-radial, therefore it 
	would require full (i.e. 3D) hydrodynamics. We can pose the above 
	question only in the case of RRd stars and double-mode Cepheids, 
	since these variables are thought to be pure radial pulsators and we 
	have successful nonlinear modelling available on these variables.

	Data on the nonlinear/linear period difference for asymptotic 
	hydrodynamical models are nearly non-existent. In an early 
	assessment \citep{kovacs01book} we used the somewhat 
	limited and preliminary result obtained by \citet{kollath01book}. 
	The nonlinear periods were longer than the linear ones in the order 
	of $\sim\,0.001$~days. The amounts of the period shifts were such that they 
	resulted in a {\em decrease} of {\em less} than $0.002$ in the period ratio. 
	An ongoing, more extended study by Szab\'o et al. (to be published) 
	basically confirms the above result. 
	As it is detailed in Appendix \ref{appendix:lna-nl}, 
	we generated another set of LNA models 
	by artificially decreasing the observed periods of BS\,Com by the 
	amount predicted from the models of Szab\'o et al. After repeating the 
	same computation as for the LNA periods, we ended up with stellar 
	parameters that were remarkably close to the ones obtained with the 
	LNA assumption. The solution curves (age vs. stellar parameters) exhibit 
	small overall shifts that result in differences between the parameters 
	obtained with the LNA and modified LNA solutions. The parameter shifts are
	approximately as follows: 
	$+0.01$, $+0.006$, $+10$\,K, $-0.08$, for $M/M_{\odot}$, $\log(L/L_{\odot})$,
	$T_{\rm eff}$ and 
	[Fe/H], respectively, in the sense of modified LNA minus non-modified 
	LNA. We conclude that considering the period shifts predicted by 
	the currently available nonlinear simulations does not effect significantly 
	the agreement with the independently estimated $T_{\rm eff}$ and [Fe/H].


\section{Conclusions}

	We performed a pulsational/evolutionary analysis on one of the most 
	extensively observed Galactic double-mode RR~Lyrae (RRd) star BS\,Com. 
	Our earlier analysis incorporating all of our measurements in the 
	$BV(RI)_{\rm C}$ colours has shown that the light variation does not contain any 
	other components except for those of the fundamental and first overtone
	modes together with their linear combinations \citep{dekany07an}. 
	Therefore, the presence of possible non-radial components is excluded 
	above the $\sim$\,mmag level. (This may not be the case for AQ~Leo, where 
	the analysis of the high precision data from the MOST satellite suggests 
	the presence of two possibly non-radial components at the sub-mmag 
	level -- see \citealt{gruberbauer07mnras}.)
	Based on the result of the frequency analysis, 
	we decided to derive the physical parameters of BS\,Com using the 
	assumption of radial double-mode pulsation.   
	
	Earlier studies on double-mode variables were mostly based on the 
	period ratio (Petersen) diagram, which is an effective (but not 
	complete) representation of the main (mass and heavy element) 
	dependences of the periods \citep[see e.g.][]{popielski2000aca}. 
	It is also possible to employ other available pieces of information, 
	such as colour or overall cluster metallicity \citep[e.g.][]{kovacs2000smc}. 
	In this paper we followed a {\em purely} theoretical approach, where 
	we derived all basic stellar parameters by matching the evolutionary 
	and pulsational properties. In the solutions age remains a free
	parameter.
	
	It is important to confront the above {\em theoretical} solution 
	with other independent pieces of information. For this goal, we 
	utilised the observed $BVI_{\rm C}$ colours in deriving metal abundance and 
	effective temperature with the help of updated solar-scaled stellar 
	atmosphere models by \citet{castelli99aap}. We tested the reliability 
	of the method through the comparison of the published spectroscopic 
	abundances of 21 Galactic field fundamental mode RR~Lyrae (RRab) stars.
	The standard deviation of the residuals between the spectroscopic 
	and photometric metallicities is $0.15$~dex, which we consider as a 
	good support for the reliability of the photometric method based on 
	the $BVI_{\rm C}$ colours. By applying this method on BS\,Com, we get 
	[Fe/H]$ = -1.58\pm 0.11$, $T_{\rm eff} = 6842\pm 10$\,K,
	where the errors are purely statistical ones.
	
	Taking into consideration the above independently estimated parameters,
	we can most probably exclude the late evolutionary
	stages after the core helium exhaustion. 
	This is because the high temperature required by the solution is 
	incompatible with the temperature computed from the observed colours.
	Therefore, in the acceptable set of solutions the age
	spans only an approximately 60~Myr range from the zero age horizontal
	branch till slightly after the start of the final redward evolution.
	Because we do not have independent information about the evolutionary status,
	all of these solutions are viable. However, the ranges covered by the parameters 
	in the possible time interval are remarkably narrow. Furthermore, assuming 
	uniform age distribution, most of the values are clumped in an even 
	narrower regime at the earlier evolutionary stages. 
	By considering this age ambiguity, we get $M / M_{\odot} = 0.698 \pm 0.004$,
	$\log(L/L_{\odot}) = 1.712 \pm 0.005$, $T_{\rm eff} = 6840 \pm 14$\,K, 
	[Fe/H]=$-1.67 \pm 0.01$ where the errors are $1\,\sigma$ ranges due to the
	above age interval.
	
	It is interesting to compare the above values with the compilation
	of the average physical parameters of the RRd stars found in various
	stellar systems. In Table~2 of the paper by \citet{kovacs2001aap} we summarised the
	result for four systems, where the parameter determinations were performed
	without the use of evolutionary models. There are IC\,4499 and LMC that have average
	metallicities similar to that of BS\,Com. Omitting LMC, due to its very wide population
	range, we see that the average RRd parameters in IC\,4499
	($T_{\rm eff} = 6760$\,K, $\log L / L_{\odot} = 1.708$, $M / M_{\odot} = 0.755$)
	are remarkably similar to those of BS\,Com.
	The largest difference occurs in the mass, but this is due to
	the slightly longer period of BS\,Com than the average value of the RRd stars
	in IC\,4499. Since the other stellar parameters are nearly the same, this period
	difference results in a larger average mass for the cluster stars.
	
	Another indirect check of the consistency of our theoretical 
	parameters of BS\,Com is the computation of the zero point of the 
	period-luminosity-colour (PLC) relation of \citet{kw01aap}. 
	Although this relation was derived from fundamental mode RR~Lyrae 
	stars, because of its apparent generality, we use it with the 
	period of the fundamental mode of BS\,Com. By using 
	$M_{\rm bol,\odot} = 4.75$, a bolometric correction interpolated 
	to the appropriate stellar parameters of ${\rm BC} = -0.15$
	\citep[see][]{castelli99aap} and a reddening value given by
	\citet[][see Table~3]{schlegel98apj},
	we get $-1.02$ for the PLC zero point.
	In comparing this value with the one published by \citet{kovacs03mnras}, we 
	see that the above value is larger by $0.06$, yielding lower 
	distance modulus for the LMC by the same amount (we recall that 
	the LMC distance modulus with the zero point of \citet{kovacs03mnras}
	comes out to be $18.55$). We think that the above result confirms 
	the consistency of the derived physical parameters, especially if 
	we add the ambiguity due to the yet to be shown fit of the RRd 
	variables to the PLC relation derived from RRab stars. 
	
	We have also tested the effect of a possible 
	period shift due to convection and nonlinearity. Based on the 
	extensive set of models by Szab\'o et al. (in preparation), our conclusion 
	is that the effect in the derived parameters is relatively small
	($\sim\,1.5\,\%$).
	Once this correction is proved to be important by future 
	studies, taking it into consideration is fairly easy, since the 
	overall period shift seems to be almost independent of the particular 
	model parameters. Therefore, there is no need to run specific 
	nonlinear models in each case but it is quite enough to fit the 
	relatively inexpensive linear models to the modified periods based 
	on the above general corrections.
	
	The remarkably small ambiguity of the derived stellar parameters 
	from the purely theoretical method employed in this paper brings 
	up several intriguing questions. One of these is the compatibility 
	of the theoretical abundance and temperature and the ones obtained 
	by spectroscopic and various semi-empirical methods (e.g. by the 
	widely used infrared flux method).
	Accuracies such as $1\%$ (or better) are quoted for the temperature
	in the current papers on this topic \citep*[e.g.][]{masana06aap}.  
	Unfortunately, for the chemical composition we still lack this 
	level of accuracy. Most of the abundances available for RR~Lyrae 
	stars are based on some calibration of low-dispersion spectra by 
	a low number of high-dispersion ones. Double-mode stars in globular 
	clusters 
	are obvious targets for the above-type of studies, due to the expected 
	chemical homogeneity and same distance. Last, but not least, we 
	note the need of putting more stringent constraints on the nonlinear 
	hydrodynamical models. Our current understanding is that double-mode 
	behaviour is caused by the non-resonant interaction of two normal 
	modes with a delicate level of physical dissipation (understood as 
	the result of convection). Although the global observational and 
	theoretical properties are in reasonable agreement, the accuracy 
	of the presently available stellar parameters is insufficient to lend 
	further support to these models. Considering the general astrophysical 
	importance of convection and, in particular, the role of that in double-mode 
	pulsation, it is clear that pinning down the physical parameters is 
	of great significance.  
	

\section*{Acknowledgements}
	We are very grateful to Zolt\'an Koll\'ath for lending his 
	opacity interpolation code.
	Thanks are also due to L\'aszl\'o Szabados for his help in the
	language correction of the paper.
	R.\,Sz. is grateful to the Hungarian NIIF
	Supercomputing Center for providing resources for the nonlinear
	model computations (project No. 1107).
	We are also grateful to Csaba Kom\'aromi for his valuable help
	in constructing the aperture stop to obtain standard magnitudes of BS\,Com.
	This work has been supported by the Hungarian Research Foundation
	(OTKA) grants K-60750 (to G.\,K.) and K-68626 (to J.\,J.).




\appendix
		
\section{Fourier decomposition of BS~Comae}
\label{appendix:decomp}
	Table~\ref{tab:fourier} shows the frequencies, amplitudes and phases
	of the 15-term Fourier sum fitted to the light curves of BS\,Com.
	Phases refer to sine-term decomposition with an epoch $t_0 = 2453431.0$.
	The Fourier-sums fit the $B, V, R_C, I_{\mathrm{C}}$ data with
	$0.0116$, $0.0097$, $0.0091$, $0.0098$ mag rms scatters, respectively.
	These values are in the appropriate proximity of the formal errors of the
	individual photometric data points as derived from the IRAF package.
	To determine the errors of the fitted amplitudes and phases,
	we performed Monte Carlo simulations of the light curves.
	Synthetic data were generated by adding independent Gaussian random
	noise to the Fourier solution, with $\sigma$ made equal to the rms
	scatter of the residual data.
	Errors in Table~\ref{tab:fourier} are the standard deviations
	of the Fourier parameters obtained from 100 independent realisations
	of the light curve in each colour.

	\begin{table*}
	\begin{minipage}{140mm}
	\caption{Fourier decompositions of the $B, V, R_{\rm C}, I_{\rm C}$ light curves of BS\,Com.}
	\label{tab:fourier}
	\begin{tabular}{ccrrrrr}	
	\hline
	\hline
	\multicolumn{2}{c}{Identification} & Frequency &
	\multicolumn{1}{c}{$A_B$} & \multicolumn{1}{c}{$A_V$} &
	\multicolumn{1}{c}{$A_{R_{\mathrm{C}}}$} &
	\multicolumn{1}{c}{$A_{I_{\mathrm{C}}}$} \\
	\hline
	$f_0$  & --     & $2.04955$  & $164.26 \pm 0.77$ & $127.05 \pm 0.47$ & $101.18 \pm 0.45$ & $79.63  \pm 0.51$ \\
	$2f_0$ & --     & $4.09910$  & $26.51  \pm 0.70$ & $20.87  \pm 0.47$ & $17.26  \pm 0.46$ & $14.88  \pm 0.48$ \\
	$3f_0$ & --     & $6.14865$  & $4.43   \pm 0.78$ & $1.82   \pm 0.47$ & $1.44   \pm 0.45$ & $2.48   \pm 0.47$ \\
	--     & $f_1$  & $2.75432$  & $259.43 \pm 0.76$ & $201.66 \pm 0.46$ & $161.36 \pm 0.42$ & $123.06 \pm 0.51$ \\
	--     & $2f_1$ & $5.50864$  & $35.55  \pm 0.83$ & $29.08  \pm 0.49$ & $22.76  \pm 0.50$ & $17.42  \pm 0.47$ \\
	--     & $3f_1$ & $8.26296$  & $15.57  \pm 0.68$ & $12.48  \pm 0.46$ & $9.87   \pm 0.49$ & $6.86   \pm 0.44$ \\
	--     & $4f_1$ & $11.01728$ & $3.75   \pm 0.65$ & $3.21   \pm 0.42$ & $2.94   \pm 0.45$ & $2.48   \pm 0.47$ \\
	$f_0$  & $f_1$  & $4.80387$  & $62.50  \pm 0.65$ & $50.23  \pm 0.48$ & $40.87  \pm 0.54$ & $32.82  \pm 0.54$ \\
	$f_0$  & $-f_1$ & $0.70477$  & $36.19  \pm 0.68$ & $29.65  \pm 0.43$ & $24.98  \pm 0.49$ & $18.10  \pm 0.53$ \\
	$2f_0$ & $f_1$  & $6.85342$  & $11.59  \pm 0.78$ & $9.75   \pm 0.51$ & $8.54   \pm 0.42$ & $6.86   \pm 0.52$ \\
	$f_0$  & $2f_1$ & $7.55819$  & $21.67  \pm 0.74$ & $20.31  \pm 0.47$ & $15.30  \pm 0.49$ & $11.38  \pm 0.57$ \\
	$2f_0$ & $2f_1$ & $9.60774$  & $10.85  \pm 0.77$ & $6.08   \pm 0.46$ & $4.99   \pm 0.41$ & $3.14   \pm 0.45$ \\
	$f_0$  & $3f_1$ & $10.31251$ & $8.04   \pm 0.79$ & $6.46   \pm 0.46$ & $4.28   \pm 0.42$ & $3.61   \pm 0.50$ \\
	$f_0$  & $-3f_1$& $6.21341$  & $5.09   \pm 0.83$ & $2.63   \pm 0.50$ & $3.52   \pm 0.47$ & $2.24   \pm 0.50$ \\
	$f_0$  & $4f_1$ & $13.06683$ & $3.35   \pm 0.56$ & $2.56   \pm 0.45$ & $1.78   \pm 0.43$ & $1.46   \pm 0.43$ \\
	\hline
	& & & \multicolumn{1}{c}{$\Phi_B$} & \multicolumn{1}{c}{$\Phi_V$} &
	\multicolumn{1}{c}{$\Phi_{R_{\rm C}}$} & \multicolumn{1}{c}{$\Phi_{I_{\rm C}}$} \\
	\hline
	--     & $f_1$  & $2.754320$  & $3.121 \pm 0.004$ & $3.080 \pm 0.004$ & $3.034 \pm 0.005$ & $2.960 \pm 0.006$ \\
	--     & $2f_1$ & $5.508640$  & $2.255 \pm 0.030$ & $2.199 \pm 0.019$ & $2.304 \pm 0.026$ & $2.210 \pm 0.032$ \\
	--     & $3f_1$ & $8.262960$  & $1.023 \pm 0.176$ & $1.562 \pm 0.273$ & $2.202 \pm 0.355$ & $1.433 \pm 0.260$ \\
	--     & $4f_1$ & $11.017280$ & $2.216 \pm 0.003$ & $2.200 \pm 0.003$ & $2.172 \pm 0.003$ & $2.104 \pm 0.004$ \\
	$f_0$  & --     & $2.049550$  & $1.560 \pm 0.020$ & $1.534 \pm 0.016$ & $1.502 \pm 0.019$ & $1.569 \pm 0.030$ \\
	$2f_0$ & --     & $4.099100$  & $0.438 \pm 0.046$ & $0.433 \pm 0.035$ & $0.426 \pm 0.048$ & $0.457 \pm 0.074$ \\
	$3f_0$ & --     & $6.148650$  & $5.553 \pm 0.180$ & $5.992 \pm 0.133$ & $5.746 \pm 0.137$ & $5.727 \pm 0.169$ \\
	$f_0$  & $f_1$  & $4.803870$  & $1.564 \pm 0.014$ & $1.576 \pm 0.008$ & $1.602 \pm 0.013$ & $1.542 \pm 0.014$ \\
	$f_0$  & $-f_1$ & $0.704770$  & $4.576 \pm 0.022$ & $4.616 \pm 0.015$ & $4.657 \pm 0.018$ & $4.682 \pm 0.024$ \\
	$2f_0$ & $f_1$  & $6.853420$  & $0.875 \pm 0.054$ & $0.829 \pm 0.046$ & $0.931 \pm 0.059$ & $0.793 \pm 0.066$ \\
	$f_0$  & $2f_1$ & $7.558190$  & $0.234 \pm 0.035$ & $0.251 \pm 0.023$ & $0.250 \pm 0.027$ & $0.316 \pm 0.045$ \\
	$2f_0$ & $2f_1$ & $9.607740$  & $0.319 \pm 0.071$ & $0.159 \pm 0.077$ & $0.113 \pm 0.071$ & $0.178 \pm 0.135$ \\
	$f_0$  & $3f_1$ & $10.312510$ & $0.698 \pm 0.097$ & $0.359 \pm 0.082$ & $0.102 \pm 0.109$ & $6.266 \pm 0.114$ \\
	$f_0$  & $-3f_1$& $6.213410$  & $4.988 \pm 0.162$ & $5.180 \pm 0.199$ & $5.342 \pm 0.145$ & $5.418 \pm 0.236$ \\
	$f_0$  & $4f_1$ & $13.066830$ & $5.980 \pm 0.164$ & $5.303 \pm 0.170$ & $5.341 \pm 0.232$ & $4.824 \pm 0.358$ \\
	\hline
	\end{tabular}
	\begin{flushleft}
 	{\footnotesize \underline{Notes:}
 	The light curves are represented in the from of 
 	$A_0 + A_1\cdot\sin(2\pi f (t - t_0) + \Phi_1) + \dots$\\
 	with $t_0 = 2453431.0$. Frequency is given in d$^{-1}$, amplitudes and phases are in 
 	mmag and radians, respectively. We show $1\sigma$ formal statistical errors.}
 	\end{flushleft}
	\end{minipage}
	\end{table*}

\section{The sample for testing the $\bld{BVI_C}~-~\bld{[F\lowercase{e}/H],T_{\lowercase{eff}}}$ relation}
\label{appendix:teststars}

	Here we present the parameters of the 24 RRab used in Sect.~\ref{sect:abundances} 
	to test the method of metallicity and temperature estimation from the $BVI_{\rm C}$ 
	colours and fundamental periods. Table~\ref{tab:teststars} shows fundamental periods, 
	colour indices, interstellar reddenings and spectroscopic [Fe/H]$_{\mathrm{sp}}$ 
	values taken from the literature.
	We relied mainly on the compilations by \citet{kj97aap} for colour indices and 
	on \citet{jk96} for [Fe/H], except for three stars. For the abundances of UU\,Cet 
	and V440\,Sgr we refer to \citet{clementini95aj} and for RV\,Phe to \citet{jones73apjs}. 
	Reddenings were taken from \citet{blanco92aj}, except for BB\,Pup for which we used 
	the E($B-V$) value given by \citet{schlegel98apj}. The derived [Fe/H]$_{\mathrm{ph}}$ 
	photometric abundances for each star are listed in the last column of 
	Table~\ref{tab:teststars}. For X\,Ari, SS\,Leo, and VY\,Ser, we get discrepant 
	[Fe/H]$_{\mathrm{ph}}$ values. These might be due to systematic errors
	in their colour indices, which have a particularly large effect on the estimated 
	[Fe/H]$_{\mathrm{ph}}$ at lower metallicities (see Kov\'acs 2008). We also note that 
	SS\,Leo and VY\,Ser have also been met as outliers in various earlier studies 
	\citep[e.g.][]{kovacskanbur98, kw01aap, kovacs03mnras}.
	
	\begin{table*}
	\begin{minipage}{140mm}
	\caption{Properties of the test RRab stars}
	\label{tab:teststars}
	\begin{tabular}{lcccccc}
	\hline
	\hline
	\multicolumn{1}{c}{Star} & $P_{\mathrm FU}$ & $B-V$ & $V-I_{\mathrm{C}}$ & $\mathrm{E}(B-V)$ & [Fe/H]$_{\mathrm{sp}}$ & [Fe/H]$_{\mathrm{ph}}$ \\
	\hline
	SW\,And   & $0.4422659$ & $0.434$ & $0.544$ & $0.09$ &  $-0.06$ & $-0.22$ \\
	WY\,Ant   & $0.5743365$ & $0.384$ & $0.549$ & $0.05$ &  $-1.39$ & $-1.13$ \\
	X\,Ari    & $0.6511597$ & $0.490$ & $0.724$ & $0.16$ &  $-2.10$ & $-2.50$ \\
	RR\,Cet   & $0.5530288$ & $0.371$ & $0.537$ & $0.05$ &  $-1.29$ & $-1.28$ \\
	UU\,Cet   & $0.6053409$ & $0.390$ & $0.567$ & $0.01$ &  $-1.38$ & $-1.33$ \\
	W\,Crt    & $0.4120139$ & $0.370$ & $0.466$ & $0.09$ &  $-0.45$ & $-0.22$ \\
	DX\,Del   & $0.4726182$ & $0.454$ & $0.571$ & $0.09$ &  $-0.32$ & $-0.26$ \\
	SU\,Dra   & $0.6604200$ & $0.348$ & $0.519$ & $0.02$ &  $-1.56$ & $-1.50$ \\
	SW\,Dra   & $0.5696710$ & $0.369$ & $0.520$ & $0.01$ &  $-0.95$ & $-0.93$ \\
	RX\,Eri   & $0.5872475$ & $0.415$ & $0.585$ & $0.10$ &  $-1.07$ & $-1.13$ \\
	RR\,Gem   & $0.3973082$ & $0.406$ & $0.511$ & $0.11$ &  $-0.14$ & $-0.23$ \\
	RR\,Leo   & $0.4523926$ & $0.334$ & $0.486$ & $0.03$ &  $-1.30$ & $-1.23$ \\
	SS\,Leo   & $0.6263438$ & $0.350$ & $0.488$ & $0.02$ & $-1.56$ & $-0.82$ \\
	TT\,Lyn   & $0.5974332$ & $0.381$ & $0.550$ & $0.05$ &  $-1.50$ & $-1.26$ \\
	V445\,Oph & $0.3970232$ & $0.620$ & $0.772$ & $0.29$ &  $+0.01$ & $-0.18$ \\
	AV\,Peg   & $0.3903760$ & $0.424$ & $0.528$ & $0.10$ &  $+0.08$ & $-0.17$ \\
	AR\,Per   & $0.4255489$ & $0.676$ & $0.853$ & $0.35$ &  $-0.14$ & $-0.31$ \\
	RV\,Phe   & $0.5964182$ & $0.370$ & $0.535$ & $0.03$ &  $-1.50$ & $-1.19$ \\
	BB\,Pup   & $0.4805468$ & $0.459$ & $0.582$ & $0.11$ &  $-0.35$ & $-0.47$ \\
	V440\,Sgr & $0.4774788$ & $0.404$ & $0.568$ & $0.09$ &  $-1.14$ & $-1.09$ \\
	VY\,Ser   & $0.7140962$ & $0.377$ & $0.579$ & $0.02$ & $-1.71$ & $-2.33$ \\
	W\,Tuc    & $0.6422370$ & $0.329$ & $0.490$ & $0.01$ &  $-1.37$ & $-1.39$ \\
	TU\,UMa   & $0.5576570$ & $0.364$ & $0.522$ & $0.05$ &  $-1.15$ & $-1.15$ \\
	UU\,Vir   & $0.4756062$ & $0.349$ & $0.475$ & $0.02$ &  $-0.60$ & $-0.66$ \\
	\hline
	\end{tabular}
  	\begin{flushleft}
  	{\footnotesize \underline{Note:} Periods are in [days], colour indices and 
  	reddenings are in [mag]. See Appendix~\ref{appendix:teststars} for references 
  	on the sources of data.}
  	\end{flushleft}
 	\end{minipage}
	\end{table*}
	
\section{Effect of nonlinearity}
\label{appendix:lna-nl}
	
	\begin{table}
	\label{testlna}
	\caption{Pulsation models matching the modified periods of BS\,Com}
	\begin{flushleft}                                                     
	\begin{tabular}{cccccc}                                         
	\hline\hline
	$T_{\rm eff}$ & $M$ & $\log L$ & $R$ & $\log g$ & $\log \rho$  \\
	\hline
	&&\multicolumn{2}{c}{$X=0.7547$} & {$Z=0.0003$} \\
	\hline
		6700 & 0.672 & 1.662 & 5.026 & 2.862 & $-2.128$ \\
		6750 & 0.679 & 1.679 & 5.050 & 2.863 & $-2.129$ \\ 
		6800 & 0.689 & 1.697 & 5.080 & 2.864 & $-2.131$ \\
		6850 & 0.697 & 1.714 & 5.105 & 2.865 & $-2.132$ \\
		6900 & 0.707 & 1.732 & 5.137 & 2.866 & $-2.134$ \\
	\hline
	&&\multicolumn{2}{c}{$X=0.7546$} & {$Z=0.0004$} \\
	\hline
		6700 & 0.705 & 1.677 & 5.114 & 2.868 & $-2.130$ \\
		6750 & 0.715 & 1.695 & 5.144 & 2.869 & $-2.131$ \\
		6800 & 0.724 & 1.713 & 5.172 & 2.870 & $-2.133$ \\
		6850 & 0.733 & 1.730 & 5.200 & 2.871 & $-2.134$ \\
		6900 & 0.745 & 1.748 & 5.232 & 2.872 & $-2.135$ \\
	
	\hline
	&&\multicolumn{2}{c}{$X=0.7534$} & {$Z=0.0006$} \\
	\hline
		6700 & 0.772 & 1.706 & 5.287 & 2.879 & $-2.133$ \\
		6750 & 0.786 & 1.725 & 5.325 & 2.881 & $-2.135$ \\
		6800 & 0.798 & 1.743 & 5.356 & 2.882 & $-2.136$ \\
		6850 & 0.810 & 1.761 & 5.389 & 2.883 & $-2.138$ \\
		6900 & 0.820 & 1.778 & 5.416 & 2.884 & $-2.139$ \\
	\hline
	&&\multicolumn{2}{c}{$X=0.7530$} & {$Z=0.0010$} \\
	\hline
		6700 & 0.919 & 1.761 & 5.633 & 2.900 & $-2.140$ \\
		6750 & 0.930 & 1.778 & 5.660 & 2.901 & $-2.141$ \\
		6800 & 0.944 & 1.796 & 5.694 & 2.902 & $-2.142$ \\
		6850 & 0.959 & 1.814 & 5.728 & 2.903 & $-2.144$ \\
		6900 & 0.971 & 1.831 & 5.757 & 2.904 & $-2.145$ \\
	\hline                                                                
	\end{tabular}                                                         
	\end{flushleft}                                                       
	{\footnotesize\underline {Notes:} 
	$T_{\rm eff}$ is given in [K], $M$, $\log L$ and $R$ are measured in 
	solar units, $g$ and $\rho$ are in [CGS]. Observed periods match 
	within $10^{-5}$~d. The modified periods are $P_0=0.486448$ and 
	$P_1=0.362340$~days.}                                          
	\end{table}
	
	In Szab\'o et al. (in preparation) we examine the
	period shifts between the asymptotic nonlinear 
	period of the convective models and those of the purely radiative linear
	non-adiabatic models (i.e. the ones we use in this paper). The results 
	are based on many different models, including various combinations of 
	physical and model construction parameters (notably depth of the inner 
	boundary and number of mass shells constituting the stellar model). 
	All models show that both modes have longer periods in the asymptotic 
	double-mode regime than those in the LNA approximation. The amount 
	of period increase is smaller than $0.001$~days. The shifts lead to
	period ratio values that are lower by up to $0.0015$ than the LNA ones.
	For the overall relative period shifts we get:
	\begin{eqnarray}
	P_{\rm LNA}(FU) / P_{\rm NL}(FU) &=& 0.997 \\
	P_{\rm LNA}(FO) / P_{\rm NL}(FO) &=& 0.998~.
	\end{eqnarray}
	Assuming that the observed periods of BS\,Com correspond to the nonlinear 
	asymptotic values, the predicted {\em modified} LNA values are as follows: 
	\begin{eqnarray}
	P_0 (\rm{LNA}, mod.)     & = & 0.486448 \\
	P_1 (\rm{LNA}, mod.)     & = & 0.362340 \\
	P_1/P_0 (\rm{LNA}, mod.) & = & 0.744869~.
	\end{eqnarray}
	With these periods we performed the same type of LNA survey 
	as described in Sect.~5 for the observed periods. The result is shown 
	in Table~C1. By comparing the items at the same chemical composition 
	and temperature with those matching the observed periods, we see that $M$
	and $\log L$ are shifted to larger values by $\sim\,0.04$ and  
	$\sim\,0.015$, respectively.
	However, in the final solution the difference between the parameters
	obtained by the predicted and directly computed LNA periods will be
	considerably smaller, due to the compensating effect of changes
	in $T_{\rm eff}$ and [Fe/H] (see Sect.~5 for additional details).

\label{lastpage}

\end{document}